\documentclass[aps,pra,twocolumn,showpacs]{revtex4}
\usepackage{times}
\usepackage{amsmath}
\usepackage{amssymb}
\usepackage{graphicx}

\newcommand{\be}{\begin{equation}}
\newcommand{\ee}{\end{equation}}
\newcommand{\bem}{\begin{multline}}
\newcommand{\eem}{\end{multline}}
\newcommand{\bea}{\begin{eqnarray}}
\newcommand{\eea}{\end{eqnarray}}
\newcommand{\bes}{\begin{subequations}}
\newcommand{\ees}{\end{subequations}}

\newcommand{\sub}[1]{{\rm #1}}

\def\ud{\mathrm{d}}
\def\kp{k_\sub{p}}
\def\op{\omega_\sub{p}}

\begin{document}

\title{Geometric resonance cooling of polarizable particles in an optical waveguide} 
\author{G.\ Szirmai}
\author{P.\ Domokos} 
\affiliation{Research Institute of Solid State Physics and Optics, Hungarian Academy of Sciences, H-1525 Budapest P.O. Box 49, Hungary}

\begin{abstract} 
In the radiation field of an optical waveguide, the Rayleigh scattering of photons is shown to result in a strongly velocity-dependent force on atoms.  The pump field, which is injected in the fundamental branch of the waveguide, is favorably scattered by a moving atom into one of the transversely excited branches of propagating modes. All fields involved are far detuned from any resonances of the atom.  For a simple polarizable particle, a linear friction force coefficient comparable to that of cavity cooling can be achieved.
\end{abstract}


\maketitle

Laser cooling of the thermal motion of atoms relies on the velocity dependence of the light force accompanying photon scattering processes. Since the Doppler effect can lead only to a fine tuning of the scattering cross section, some kind of resonant enhancement must be involved in the light-matter interaction. In the simplest case of ``conventional methods'', one makes use  of atomic resonances driven by quasi-resonant laser sources. An alternative approach, so-called  \emph{cavity cooling} has been developed \cite{horak97,vuletic00} and demonstrated \cite{maunz04,nussmann05a} recently. In this cooling method, the atom is coupled to a discrete mode of an optical resonator.  Although the bare atomic transition frequencies can be very far from the external driving frequency, this latter can be close to a cavity eigenfrequency and thus to a polariton resonance of the coupled atom-field system \cite{domokos04}.  The virtue of this cooling scheme is that the atom behaves simply as a polarizable particle, i.e., its initial and final electronic states match. It was shown that the mechanism imparting velocity sensitivity to the optical force is a two-photon Doppler-effect \cite{vuletic01,murr:253001}.

In this Letter we show that optical cooling can be based on \emph{geometric resonances} occurring in waveguides. Here the field is confined spatially only in two directions, and propagates along the third one. The spectrum of the radiation modes of a waveguide is continuous. Unlike in cavities, as the photons are not multiply recycled by reflections, the field cannot be described by a single or by a few degrees of freedom. Therefore, even the joint atom-field system does not manifest discrete resonances, being associated with an excited state of a given mode, which would be suitable for cooling.  However, there is a discrete set of continuous branches, each of them belonging to a given spatial function in the transverse directions and to a given polarization. There are  discontinuities, a signature of the geometric resonance, at the threshold frequency of the branches, that is, when a new transversally excited mode function fits in the waveguide cross section. 

It is obvious that reducing the transverse size of the guided modes increases their coupling to an atomic dipole. For example,  a significant enhancement of the optical depth of rubidium vapour has been observed in a photonic-band-gap waveguide \cite{ghosh:023603}. Analogously, there is an enhanced  photon scattering into the waveguide modes  as well \cite{domokos02c,kien:013819}, which implies strong optical dipole forces \cite{kien:033412}. However, this effect equally applies to any mode in the continuum irrespectively to its spectral distance from the discontinuities. One has to distinguish the effects due to the threshold of excited mode branches. In particular, it gives rise to a singularity of the dipole force in a waveguide with perfectly conducting walls \cite{gomez.86.4275}. Concerning cooling, there are two further points to be considered: (i) what is the regularized form of the dipole force at threshold when the waveguide is lossy; \, (ii) how does it depend on the velocity. These problems are  solved in this Letter. 

\begin{figure}[htbp]
\includegraphics[width=0.8\columnwidth]{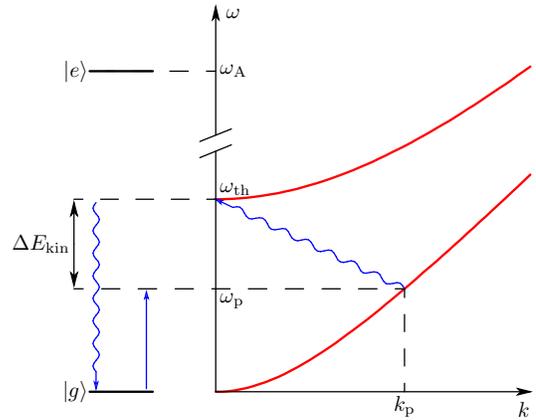}
 \caption{(Color online) Schematic representation of the Rayleigh scattering responsible for the cooling mechanism. On the right: the dependence of the waveguide mode frequency $\omega$ on the longitudinal wavenumber  $k$. The dispersion relation of the waveguide exhibits a discrete set of continuous branches. The atom absorbs a photon with frequency $\op$, wavenumber $\kp$ from the pumped mode of the fundamental branch and emits one into the excited branch. On the left: the atomic levels, the resonance frequency $\omega_A$ is very far from $\omega_p$.}
 \label{fig:scheme}
\end{figure}

The cooling mechanism, schematically shown in Fig.\ \ref{fig:scheme}, involves only Rayleigh scattering in the far-detuned limit, thus the particle remains effectively in its ground state. 
Photons are scattered from the pumped mode of the fundamental branch into other modes. The fundamental branch in the vicinity of the pump frequency $\omega_p$ has a free-space-like (linear) dispersion curve. Therefore when the photon is scattered within the fundamental branch, the mechanical effect is analogous to free-space light forces. However, there is a new channel opened for scattering: slightly above the pump frequency (the figure is not to scale), there is the onset of a new type of propagating modes. At threshold  $\omega_{\rm th}$, the corresponding field modes hardly propagate along the waveguide, i.e.\ $k\approx 0$ on this branch. The spectral density of modes is very large,  for perfect transverse confinement it is infinite.  The scattering being enhanced into this channel, the particle tends to convert the photon frequency up.  The needed energy has to be provided by the atom losing from its kinetic energy. Because of the off-resonant excitation, i.e., the pump is below the threshold, this scattering process certainly relies on the Doppler-shift and thus the accompanying force must have a significant velocity dependence. Note that similar mechanism is in the basis of the bad-cavity limit of cavity cooling \cite{vuletic01}.

The waveguide is described by a generic one-dimensional model. The dispersion relation is given by $\omega_n(k)$, where $n$ is the branch index and $k$ is the wavenumber along the propagation direction $z$.  One of the modes in the fundamental branch is pumped by an external laser source with frequency $\op$, wavenumber $\kp$.  The polarizable particle $A$ with mass $m$ is modeled by an unsaturated two-level atom with resonance frequency $\omega_A$ and width $2 \gamma$. The Hamiltonian of the system, in a frame rotating with the laser frequency  $\op$, is
\bes
\be
H=H_0 + H_\sub{int} +H_\sub{pump}
\ee
\be
H_0 = \frac{p_A^2}{2m}  + \hbar \delta_A \sigma^\dagger\sigma + \sum_n\int_{-\infty}^{\infty} \ud k\, \hbar \delta_n(k) a_{kn}^\dagger a_{kn}\,,
\ee
with detunings $\delta_A= \omega_A-\op$ and $\delta_n(k)= \omega_n(k)-\op$. The field mode operators $a, a^\dagger$ and the atomic excitation operators $\sigma, \sigma^\dagger$ are coupled via
\be
H_\sub{int}= -i \hbar \sum_n\int \ud k\,  g_{kn} \left( \sigma^\dagger a_{kn} e^{i k z_A} - h.c.\right)\,,
\ee
where the coupling constant $g_{kn}=\sqrt{\frac{\sigma_A}{{\cal A}_{kn}}\, \frac{\omega_A}{\omega_n(k)}\, \frac{\gamma c}{4 \pi}}$, $\sigma_A=3\lambda^2/2\pi$ is the atom's radiative cross section, and ${\cal A}_{kn}$ is the effective mode cross section. The atom is at $z_A$ along the propagation direction, and its position in the transverse directions is incorporated into the cross section ${\cal A}_{kn}$. The selected mode with wavenumber $\kp$ is pumped by the term
\be
H_\sub{pump} = i \hbar \left( \eta a_{\kp,0}^\dagger - \eta^* a_{\kp,0} \right)\; ,
\ee
\ees
where $\eta$ describes the pump field amplitude. Besides the interactions included in this Hamiltonian, the atom and the waveguide modes are coupled to reservoirs, composed of the unguided vacuum field modes and the material degrees of freedom of the waveguide. These couplings lead to dissipative processes:  (i) to the atomic spontaneous emission into other than the waveguide modes with a rate $2 \gamma$;\;�(ii) photon loss from the waveguide which will be taken into account by a single rate parameter $\kappa$ for all waveguide modes ($\kappa/c$ gives the loss  per unit length). In what follows, for the calculation of the mean friction force, detailed microscopic modeling of the dissipative processes is not needed and the description in terms of the phenomenological loss rates, $\gamma$ and $\kappa$, can be kept very general. We emphasize that $\kappa$ is an important parameter of the model: the physical process of waveguide loss leads to a natural reguralization of the singularity in the force at threshold. 

The atomic translational motion is adiabatically separated from the internal dynamics, this latter includes the position and velocity of the atom parametrically by the formal replacement $z_A \rightarrow z_A+v_A t$. The operators $a$ and $\sigma$ obey  Heisenberg-Langevin equations from which the stationary solution for the mean values can be easily obtained. For example, the mean excitation probability reads
\begin{subequations}
\be
 \label{eq:saturation}
\langle \sigma^\dagger \sigma \rangle = \frac{g^2 \eta^2/\kappa^2}{(\delta_A+\kp v_A+\Delta_{A})^2 + (\gamma +\Gamma_{A})^2} \;,
\ee
with waveguide induced light shift and resonance broadening
\be
\label{eq:lightshift}
 i\Delta_{A} + \Gamma_{A}= \sum_n\int \ud k\,  \frac{g_{kn}^2}{i (\delta_n(k) - (k - \kp) v_A) + \kappa} \; .
\ee
\end{subequations}
This expression has been evaluated recently for the case of a specific microstructure and for an atom at rest \cite{horak03}, and was shown to yield giant light shifts when the atomic frequency is very close to, slightly below, the onset of a new propagating mode branch. In what follows, to be conform with the simple polarizability picture, we will consider an atomic frequency very far above the pump frequency and far away from any branch threshold frequency, thus the detuning $\delta_A$ is dominant in the denominator of Eq.\ (\ref{eq:saturation}). 

The force acting on the moving atom can be derived from the quasi-stationary solutions \cite{murr:253001}, assuming that the motional state of the atom varies on a slower time scale. Here we can insert them directly into the expression of the force
\be
F = - \sum_n\int \ud k\, \hbar k \, g_{kn} \left( \sigma^\dagger a_{kn} e^{i k (z_A+v_A t)} + h.c.\right) \,,
\ee
which leads to the mean value
\begin{multline}
 \label{eq:force}
\langle F (\kp) \rangle = 2 \hbar \langle \sigma^\dagger \sigma \rangle \Biggl( \kp
\gamma 
\\ + \kappa \sum_n\int \ud k\, \frac{ (\kp- k)\, g_{kn}^2}{ (\delta_n(k) - (k-\kp) v_A )^2+\kappa^2}
\Biggr) \,.
\end{multline}
This is a general result for the radiation pressure force. Any specific waveguide geometry can be considered by inserting its characteristic dispersion relation $\delta_n(k)$. The force comprises two terms, originating from the scattering into free space modes ($F_\sub{fs}$), and into the waveguide modes, respectively. The velocity dependence of the force is due to the two-photon Doppler shift in the denominator. The force can attain unusually large values for certain velocities, since the Doppler-shift term can tune the pump frequency into resonance with the modes of an excited branch. 

\begin{figure}[htbp]
\includegraphics{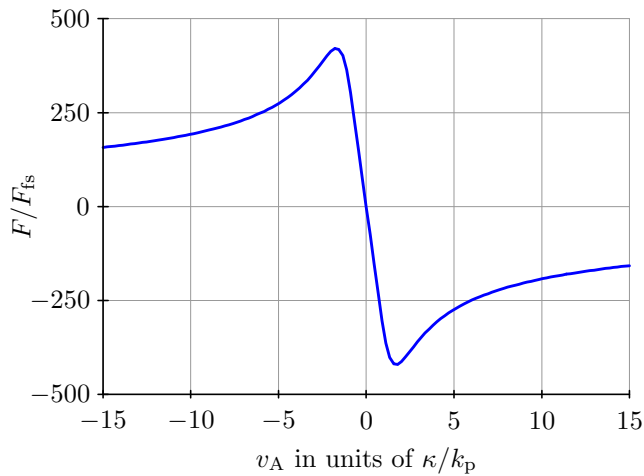}
\caption{(Color online) Velocity dependence of the force along the waveguide on an atom symmetrically illuminated from both directions. The force is normalized to the radiation pressure force $F_\sub{fs}$,  arising from scattering photons from one of the pumps into free space. System parameters are given in the text.}
\label{fig:force}
\end{figure}
Let us assume two-sided pumping, both the $\kp$ and $-\kp$ modes are driven by lasers with equal intensities but independent phases. Then the free-space contributions to the radiation pressure, $F_\sub{fs} = \pm 2 \hbar \langle \sigma^\dagger \sigma \rangle  \kp \gamma$, are balanced. It is easy to admit that, for symmetry reasons, the total force is also balanced for $v_A=0$. For moving atoms, $v_A\neq 0$, the Doppler shift in the denominator in Eq.\ (\ref{eq:force}) gives rise to a  resonance-like  velocity dependence of the force acting on the atom, as clearly manifested  in Fig.\ \ref{fig:force}. Around the origin,  the direction of the force is opposite to that of the velocity, i.e.\ there is a cooling regime. The velocity capture range is found to be on the order of $\kappa/\kp$ (this expression applies for a broad range of the waveguide loss rate $\kappa$). The velocity dependence has a linear term. The linear coefficient, i.e.\ the slope at the origin, can be identified with a characteristic friction coefficient $\beta$. 

The resonance-like functional form depicted in this figure is determined mainly by the dispersion relation of the waveguide. For this plot we considered a simple generic example: a  waveguide with a rectangular shape and conducting walls. It has the dispersion relation $\omega_n(k) =\sqrt{\omega_\sub{th,n}^2+(c k)^2}$, where $\omega_\sub{th,n}$ is the threshold of the $n$th branch and $c$ is the speed of light. In the transverse directions the modes are sinusoidal functions. The fundamental mode has no node, the first excited mode has a nodal line in the center. The atom is assumed to be a quarter of the waveguide width away from the center, where the excited mode has a maximum. The atomic detuning is chosen very large, $\delta_A=10^5\gamma$, thus the atomic resonance does not influence the results ($\gamma = 2 \times 10^7$ s$^{-1}$ for concreteness). 
For  $\kappa=10^9 $s$^{-1}$, used for Fig.\ \ref{fig:force}, the maximum of the force due to scattering into the excited branch (`waveguide force') exceeds by almost three orders of magnitude the radiation pressure due to scattering into free space at the same level of atomic saturation. The numerical value of the slope $\beta$ and the maximum of the curve strongly depends on $\kappa$. 

\begin{figure}[htbp]
\includegraphics{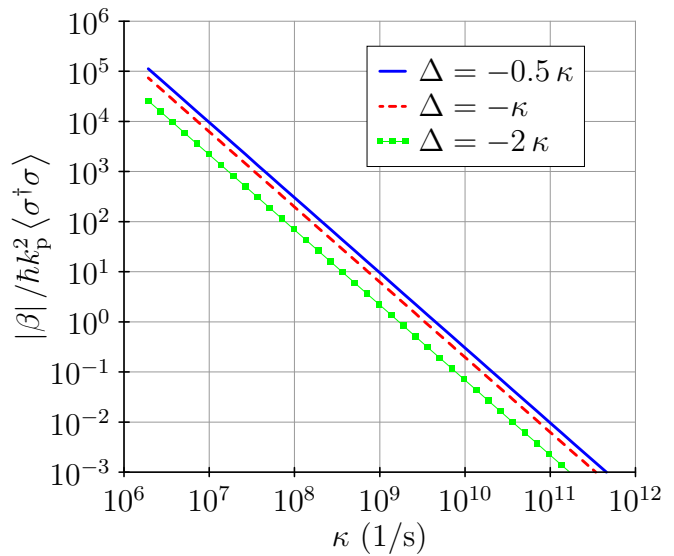}
\caption{(Color online) Power law dependence, shown in a log-log scale, of the friction coefficient $\beta$ as a function of the waveguide loss rate $\kappa$ for various detunings $\Delta = \op -\omega_\sub{th}$, where $\omega_\sub{th}$ is the threshold frequency of the first excited branch of modes.}
\label{fig:beta_kappa}
\end{figure}
The friction coefficient $\beta$ as a function of the regularization parameter $\kappa$ is plotted on a log-log scale in Fig.\ \ref{fig:beta_kappa}. $\beta$ is scaled to the maximum of the free space Doppler friction coefficient at the given excitation value $\langle \sigma^\dagger \sigma \rangle$ (this maximum would be obtained at a small detuning $\delta_A = \gamma$). Several choices of the detuning from the threshold, $\Delta = \op -\omega_\sub{th}$, are considered. One finds accurately a power law dependence $\beta \propto \kappa^{-3/2}$, where the exponent is independent of the detuning $\Delta$. Note that in cavity cooling the friction coefficient for an analogous parameter setting has a $\kappa^{-2}$ dependence, where $\kappa$ is the linewidth (HWHM) of the single cavity mode used for cooling \cite{vukics05b}.  This difference in the exponent is a consequence of the lower dimensional spatial confinement of the field in the case of waveguides.  Considering the numerical values for the friction coefficient, one finds that the cooling rate $2\beta/m$ exceeds the spontaneous emission rate $2\gamma\langle \sigma^\dagger \sigma \rangle$ for waveguide loss rates below $\kappa \approx 10^7 \ldots 10^8 $s$^{-1}$ in the case of alkali atoms. Note that this numerical value is specific to this example. In general, the cooling rate depends not only on $\kappa$, but also on the given dispersion relation $\delta_n(k)$ determined by the geometry and the transverse confinement mechanism  of the waveguide.

\begin{figure}[htbp]
\includegraphics{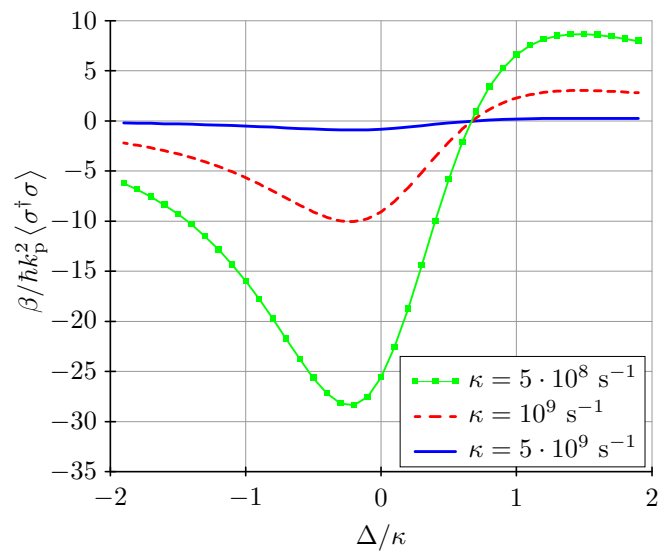}
\caption{(Color online) Resonant behaviour of the friction coefficient as a function of the pump detuning from the threshold $\Delta=\op-\omega_\sub{th}$.}
\label{fig:beta_delta}
\end{figure}
The large friction coefficient is due to the (geometric) resonant enhancement of the mode density at the threshold frequency $\omega_{\rm th}$. This is evidenced by the strong  dependence of $\beta$ on the detuning $\Delta=\op-\omega_\sub{th}$, as shown in Fig.\ \ref{fig:beta_delta}. As long as the atomic resonance $\omega_A$ is very far, the plotted curves for various $\kappa$ have similar forms, only the magnitudes are different.  The optimum detuning corresponds to a pump frequency, as expected, slightly below the threshold. However, due to the finite width of the threshold, even a pump frequency above threshold can lead to cooling force. 

The issue of coupling cold atoms to the radiation field sustained by an optical waveguide appears in various contexts. For example, hollow optical glass fibers were used to guide atoms over long distances \cite{noh02}, especially, employing red detuned light field filling out the hollow core \cite{olshanii93,PhysRevLett.75.3253}. Substrate-based atom waveguide can also be realized by using guided two-color evanescent light fields \cite{barnett:023608}. Very recently, the coupling of atomic dipoles to the evanescent field of tapered optical fibers has been demonstrated \cite{nayak06,sague07}. This system has a significant geometrical flexibility with respect to cavity-setups, which can be a major advantage, for example, in replacing cavity cooling of trapped atoms \cite{zippilli05a}. We showed that a similarly large viscous force on moving atoms can be obtained by a proper tuning of the pump field into the vicinity of a branch threshold. The proposed cooling scheme relies on Rayleigh scattering, thereby keeping the atom in the electronic ground state, which was previously known only for cavity cooling.  Finally, the present results imply that the unusually strong optical interaction between particles in a quasi-one-dimensional geometry \cite{gomez-medina:243602} is also expected to get velocity-dependence. 

We acknowledge funding from the National Scientific Fund of Hungary (NF68736, T043079 and T049234).


\end{document}